\documentclass[epj]{svjour}
\usepackage{latexsym}
\usepackage{graphicx}
\usepackage[usenames,dvipsnames]{xcolor} 
\usepackage{amsmath,amssymb}

\begin{document}
\title{The variational method, backreactions, and the absorption probability in Wald type problems}
\author{Koray D\"{u}zta\c{s}
}                     
%

\institute{Department of Natural and Mathematical Sciences,
\"{O}zye\u{g}in University, 34794 \.{I}stanbul Turkey }

\date{Received: date / Revised version: date}
%
\abstract{We argue that the variational method in Wald type thought experiments, involves order of magnitude problems when one imposes the fact that $\delta M$ is inherently a first order quantity itself. One observes that the contribution of the second order perturbations is actually of the fourth order. Therefore backreactions have to be explicitly calculated. Here,  we re-consider the overspinning problem for Kerr-Newman black holes interacting with test fields. We calculate the backreaction effects due to the induced increase in the angular velocity of the event horizon, which brings a partial solution to the overspinning problem.  To bring an ultimate solution, we argue that the absorption probability should be taken into account in Wald type problems where black holes interact with test fields. This fundamentally alters the course of the analysis of the thought experiments. Due to the fact that a small fraction of the challenging modes is absorbed by the black holes, overspinning is prevented for both nearly extremal and extremal cases.  Some extreme cases are easily fixed by backreaction effects. The arguments do not apply to the generic overspinning by fermionic fields for which the absorption probability is positive definite.
%
} 
\maketitle
\section{Introduction}
One of the main unsolved problems in classical general relativity is the validity of the cosmic censorship conjecture due to Penrose \cite{ccc}. The conjecture aims to circumvent the problems that could arise  if a curvature singularity is allowed to be in causal contact with distant observers. This is achieved by forbidding the existence of naked singularities in a physical universe. The gravitational collapse of a massive object should end up in a black hole surrounded by an event horizon rather than a naked singularity, as prescribed by Penrose and Hawking \cite{singtheo}. 

The natural question at this stage is whether the event horizon of a black hole can be destroyed  by test bodies or fields to expose the curvature singularity lurking at the center. The possibility to destroy an event horizon was first evaluated in a thought experiment constructed by Wald~\cite{wald74}. Wald started with an extremal Kerr-Newman black hole and attempted to increase the charge and angular momentum beyond the extremal limit by sending in test bodies from infinity. It turns out that the test bodies that could potentially overcharge or overspin an extremal Kerr-Newman black hole are not absorbed by the black hole. The event horizon is stable and the smooth structure of the space-time is maintained excluding the black hole region inside the event horizon. Later, Hubeny adapted an alternative approach to Wald type problems where one starts with a nearly extremal black hole instead of an extremal one~\cite{hu}. She showed that a nearly extremal Reissner-Nordstr\"{o}m black hole can be overcharged into a naked singularity by a test body. Jacobson and Sotiriou applied an analogous analysis to show that nearly extremal Kerr black holes can be overspun by test bodies~\cite{js}. D\"{u}zta\c{s} and Semiz derived the same result for nearly extremal Kerr black holes interacting with test fields~\cite{overspin}. In these works the overspinning and overcharging of nearly extremal black holes are not quite generic, which suggests that they should be fixed by employing backreaction effects. It was argued that the self force effects can prevent the overcharging \cite{backhu} and overspinning \cite{backjs}
of nearly extremal black holes by test bodies. In a very recent work we showed that the absorption of the test fields that could overspin nearly extremal black holes is not allowed due to the increase in the angular velocity of the event horizon before the absorption of the field~\cite{kerrmog}.

In literature there exist various attempts to overspin or overcharge black holes with test bodies \cite{f1,saa,gao,siahaan,magne,dilat,higher,v1,he,wang,gim,jamil,shay,shay2,zeng}, and fields \cite{semiz,emccc,duztas,toth,natario,duztas2,mode,taub-nut,kerrsen,gwak5,gwak6,hong,yang,bai}. The effect of quantum tunnelling and particle creation has also been incorporated in Wald type problems \cite{q1,q2,q3,q4,q5,q6,q7}. In recent years, the possibility to destroy the event horizon in the asymptotically anti-de Sitter cases has also become an active field of research~\cite{btz,gwak1,gwak2,gwak3,gwak4,chen,he2}. ( For a recent review see \cite{ong})

In Wald type problems the backreaction effects are difficult to compute and most of the time the results are restricted to order of magnitude estimates. Recently Sorce and Wald designed a new type of gedanken experiment by adapting a variational approach~\cite{w2}. They derived an explicit expression for the second order effects, so one does not have to explicitly compute self force or finite size effects. Currently, the Sorce-Wald method is  widely accepted among the researchers working on Wald type problems. Recently we have also employed Sorce-Wald method to test the stability of the event horizon for Martinez, Teitelboim, Zanelli (MTZ) black holes \cite{mtz}. In our analysis, we have imposed the fact that $\delta M$ is inherently a first order quantity. We observed that, imposing this fact causes order of magnitude problems to arise in the method developed by Sorce and Wald. In section (\ref{sorcewald}), we further scrutinize the Sorce-Wald method by imposing the fact that $\delta M$ is inherently a first order quantity itself, for test bodies and fields. Unfortunately, it turns out that the order of magnitude problems do not pertain to the MTZ case.  They are also manifest in the case of Kerr-Newman black holes, for which the Sorce-Wald method was developed. We present the details in section (\ref{sorcewald_1}). 

The order of magnitude problems in the Sorce-Wald method suggest that the backreactions should be explicitly calculated. Based on this argument, we re-visit the overspinning problem for nearly extremal and extremal Kerr-Newman black holes interacting with test fields, in section (\ref{backreaction}). In a recent work we showed that extremal Kerr-Newman black holes which satisfy $J^2/M^4 <(1/3)$ can be overspun by scalar test fields \cite{generic}.  We argued that the overspinning is not quite generic and it is prone to be fixed by backreaction effects. In section (\ref{backreaction_nex}), we show that nearly extremal black holes can also be overspun and employ the backreaction effects based on Will's argument that the angular velocity of the event horizon increases before the absorption of the test field \cite{will}. The employment of this backreaction effect brings a partial solution to the problem. The destruction of the event horizon can be prevented for certain classes of nearly extremal and extremal black holes, with a sufficiently large magnitude of angular momentum. In section (\ref{backreaction_ex}), we show that the same argument applies to extremal black holes. We analytically derive the relevant magnitude of the initial value of angular momentum, for both nearly extremal and extremal cases.

The interaction of test fields with black holes is actually a scattering problem. The field is partially absorbed by the black hole and partially reflected back to infinity. The fact that only a fraction of the incoming field is absorbed by the black hole has been ignored in all the thought experiments constructed so far, including the works of this author. In section (\ref{prob}), we take the absorption probabilities of test fields into account, which fundamentally changes the course of the analysis of the problem. We show that a very small fraction of the challenging test fields are absorbed by the black hole which has no practical effect on the mass and angular momentum parameters of the space-time. In section (\ref{optimal}) we evaluate the optimal perturbations with the lowest possible energy relative to their angular momentum and charge. We show that the absorption probability is zero for the optimal perturbations. In that case, the test field is entirely reflected back to infinity. The space-time parameters remain identically the same after the interaction with the test field. In sections (\ref{prob_nex}) and (\ref{prob_ex}) we perturb nearly extremal and extremal black holes with challenging modes with frequencies slightly larger than the optimal perturbations. We show that the absorption probability is very low for the challenging modes. It turns out that most of the energy and angular momentum carried by the challenging modes is reflected back to infinity. Still there exists a set of fine-tuned parameters that seem to be capable of overspinning extremal and nearly extremal Kerr-Newman black holes. In sections (\ref{prob_nex}) and (\ref{prob_ex}), we also show that, these anomalies are remedied by backreaction effects due to the induced increase in the angular velocity of the event horizon. 

\section{Sorce-Wald method}
\label{sorcewald}
The Kerr-Newman metric describes a black hole surrounded by an event horizon provided that the spacetime parameters satisfy the main inequality
\begin{equation}
M^2 \geq a^2 + Q^2
\label{main}
\end{equation}
where $M$, $a\equiv J/M$, and $Q$ are respectively the mass, angular momentum and charge parameters of the spacetime. In Wald type problems, one starts with an extremal or nearly extremal black hole satisfying the main criterion (\ref{main}) and attempts to increase the angular momentum and/or charge parameters beyond the extremal limit, by sending in test bodies or fields from infinity. The main assumption in these thought experiments is that the interaction of the black hole with test bodies and fields does not alter the background geometry of the spacetime, but leads to perturbations in the mass, angular momentum, and charge parameters. After a sufficiently long time, the spacetime is supposed to settle down to a new Kerr-Newman solution with modified parameters. Apparently the energy, angular momentum, and the charge of the test body or field should be very small compared to the initial parameters of the spacetime so that the assumption that the background geometry in the final state is also a Kerr-Newman solution, is justified. Then, one can check if the final parameters of the spacetime represent a Kerr-Newman black hole satisfying the inequality (\ref{main}) or a naked singularity which violates it. For that purpose we prefer to define
\begin{equation}
\delta_{\rm{fin}} \equiv  M_{\rm{fin}}^2 -Q_{\rm{fin}}^2 - \frac{J_{\rm{fin}}^2}{M_{\rm{fin}}^2}
\label{second}
\end{equation}
If the contribution of the second order terms are taken into account in calculating  $\delta_{\rm{fin}}$, one should also incorporate the effect of backreactions which bring second order corrections to (\ref{second}), so that the calculation can be considered  consistent. As we mentioned in the introduction, the overcharging of Reissner-Nordstr\c{o}m black holes \cite{hu}, and the overspinning of Kerr black holes \cite{js,overspin} can be fixed by backreaction effects \cite{backhu,backjs,kerrmog}. The backreaction effects comprise finite size effects, self interaction, gravitational radiation, the effect of black hole radiation, induced increase in the angular velocity of the horizon and many more possible effects pertaining to the specific problem. The Sorce-Wald method has been developed to bring an ultimate solution to the problem of determining and calculating the backreactions~\cite{w2}. Sorce and Wald (SW) attempted obtain an expression for the full second order correction $\delta^2 M$ without having to calculate the backreaction effects explicitly. To check whether the event horizon can be destroyed SW first derive an expression for the minimum energy of the incoming test body or field so that it is absorbed by the black hole.
\begin{equation}
\delta M-\Omega_H \delta J-\Phi_H\delta Q \geq 0
\label{needham}
\end{equation}
where $\Omega_H=a/(r_+^2 +a^2)$, $\Phi_H =(Qr_+)/(r_+^2 +a^2)$, and $r_+$ is the horizon radius. The condition (\ref{needham}) is well known in black hole physics. The first derivation without assuming the validity of cosmic censorship known to this author is by Needham in 1980~\cite{needham}. The condition (\ref{needham}) determines the lowest possible energy for a given combination of angular momentum and charge that would allow the absorption of a test field. The perturbations with the lowest possible energy are referred to as the optimal perturbations. The perturbations that do not satisfy the condition (\ref{needham}) are not absorbed by the black hole. If the absorption of these perturbations was allowed, they would lead to a generic destruction of the event horizon since they carry relatively large angular momentum and charge. However, the condition (\ref{needham}) only applies to the perturbations that satisfy the null energy condition.  For fermionic fields there is no lower bound on the energy 
that would prevent the absorption of the challenging modes. In \cite{generic} we argued that the absence of the lower bound for the energy of the fermionic fields leads to a generic destruction of the event horizon.

Sorce and Wald proceed by parametrizing a nearly extremal black hole as
\begin{equation}
M^2-Q^2-(J/M)^2=M^2 \epsilon^2
\label{param}
\end{equation}
which is common in Wald type problems. The small parameter $\epsilon$ determines the closeness of the black hole to extremality. For $\epsilon \ll 1$ the black hole is very close to extremality in which case the effect of the interactions with test bodies and fields become relevant. Next, Sorce and Wald define the function:
\begin{equation}
f(\lambda)=M(\lambda)^2-Q(\lambda)^2-J(\lambda)^2/M(\lambda)^2
\label{lambda}
\end{equation}
If $f(\lambda)<0$ the inequality (\ref{main}) is violated and the event horizon cannot exist. Next $f(\lambda)$ is expanded to second order in $\lambda$
\begin{eqnarray}
f(\lambda)&=& \left(M^2-Q^2-\frac{J^2}{M^2} \right) \nonumber \\
&+& 2\lambda \left( \frac{M^4+J^2}{2M^3}\delta M -\frac{J}{M^2}\delta J -Q\delta Q \right) \nonumber \\
&+& \lambda^2 \left[ \frac{M^4+J^2}{2M^3}\delta^2 M -\frac{J}{M^2}\delta^2 J -Q\delta^2 Q +\frac{4J}{M^3}\delta J \delta M \right. \nonumber \\
&-&\left. \frac{1}{M^2}(\delta J)^2 + \left( \frac{M^4-3J^2}{M^4} \right) (\delta M)^2 - (\delta Q)^2 \right ]
\label{swmain}
\end{eqnarray}
To avoid any confusion we refer to $\delta M, \delta J, \delta Q$ as first order perturbations, and $\delta^2 M, (\delta M)^2, ... $ terms as second order perturbations. For the first order perturbations Needham's condition (\ref{needham}) implies that
\begin{eqnarray}
f(\lambda)&\geq & M^2 \epsilon^2 + \frac{2}{M^4 + J^2} \left( (J^2 -M^4) Q\delta Q -2JM^2 \delta J) \right) \lambda \epsilon \nonumber \\
&+& O(\lambda^2 , \epsilon^3 , \epsilon^2 \lambda )
\label{firstorder}
\end{eqnarray}
The equations (\ref{swmain}) and (\ref{firstorder}) above, are the equations (119) and (120) in the relevant paper of Sorce and Wald \cite{w2}. To derive (\ref{firstorder}), one imposes the condition (\ref{needham}) that the test body or field is absorbed by the black hole, and expresses $\delta M$ in terms of $\delta J$ and $\delta Q$. At this point Sorce and Wald claim that neglecting the terms of order $O(\lambda^2)$ it is possible to make $f(\lambda)<0$. This statement aims to convince the readers that the variational method reproduces the previous results that the nearly extremal black holes can be overcharged or overspun when the second order terms are neglected. The main claim of SW is that $f(\lambda)$ becomes positive again  by considering the effect of the terms that are second order in $\lambda$. Here we show that these two results cannot be obtained by the method developed by Sorce and Wald, when one does not ignore the fact that $\delta M$ is inherently a first order quantity, itself.
\subsection{Sorce-Wald method with the correct test body/field approximation}  
\label{sorcewald_1}
The small parameter $\lambda$ is introduced in (\ref{lambda}) to ensure that the variation of the function $f$ from its initial value $f(0)$ is small, in accord with the test body/field approximation. However, in (\ref{swmain}) the parameter $\lambda$ explicitly multiplies the terms $\delta M, \delta J, \delta Q$. Since $\delta M$ is inherently a first order quantity, the $\lambda \epsilon \delta M$ terms actually contribute to third order to $f(\lambda)$, while the $\lambda^2 \delta^2 M$ terms contribute to fourth order. One cannot  ignore the fact that $\delta M$ is a small quantity itself and proceed as if $\delta M \sim M$, as it was done in the derivation of Sorce and Wald. 

To clarify the fact that $\delta M$ is a small quantity itself, we let
\begin{equation}
\delta M = M \zeta
\label{testfield} 
\end{equation}
where $\zeta$ parametrises the energy of the perturbation and the fact that $\zeta \ll 1$ ensures that the  test body/field approximation is not violated.  In principle the parameters determining the magnitude of the perturbations and the closeness to extremality need not be equal. However, for numerical calculations one can let $\epsilon \sim \zeta$.  Imposing the fact that the perturbations are small themselves (\ref{firstorder}) implies that 
\begin{equation}
f(\lambda)\sim O(\epsilon^2) - O(\lambda \epsilon \zeta) 
\label{manifest1}
\end{equation}
which is valid to first order in $\lambda$. (Note that the term $(J^2-M^4)$ is negative.) It is easy to see that when one imposes the fact that the first order perturbations are of the ``first order'' themselves, it is not possible to make $f(\lambda)$ --defined by SW-- negative for the first order terms. The variational method does not reproduce the previous results due Hubeny \cite{hu}, Jacobson-Sotiriou \cite{js} and D\"{u}zta\c{s}-Semiz \cite{overspin}. For the first order perturbations, the results of SW contradict with the previous results when the fact that $\delta M$ is inherently a small quantity is taken into account.

Though it is manifest in (\ref{manifest1}) that  $f(\lambda)$ defined by SW, cannot be made negative for the first order terms,  it would be appropriate  to elaborate on this  subject  considering the fact that the SW method is widely accepted in black hole physics. For simplicity let us consider a neutral body or a field  $(\delta Q=0)$, incident on a nearly extremal black hole. For the optimal perturbations (\ref{needham}) implies that
\[
J\delta J =M \delta M (r_+^2 + a^2)
\]
Imposing the fact that $\delta M=M \zeta$ by the definition (\ref{testfield})
\begin{eqnarray}
J \delta J &=& M^2 \zeta \left[ M^2 (1+\epsilon)^2 + \frac{J^2}{M^2} \right ] \nonumber  \\
&=& (M^4 + J^2 )\zeta + O(\epsilon \zeta, \epsilon^2 \zeta)
\label{jdeltaj}
\end{eqnarray}
where we have substituted $r_+=M(1+\epsilon)$ for a nearly extremal black hole parametrized as (\ref{param}). Now we substitute the expression for $J \delta J$ derived in (\ref{jdeltaj}) to the expression for $f(\lambda)$. For the optimal perturbations one derives
\begin{equation}
f(\lambda)=M^2 \epsilon^2 - 4M^2 \lambda \epsilon \zeta - O(\lambda\epsilon^2 \zeta, \lambda\epsilon^3 \zeta )
\label{manifest2}
\end{equation}
Again it is manifest in (\ref{manifest2}) that $f(\lambda)$ defined by SW, cannot be made negative for small $\lambda$, $\epsilon$ and $\zeta$. The claim that  $f(\lambda)$ can be made negative by the terms first order in $\lambda$
requires  one  to assume that $\delta M \sim M$, which apparently contradicts the test body/field approximation. 

The main claim of SW is that the negativeness of $f(\lambda)$ can be fixed by the contribution of the terms that are second order in $\lambda$. Though the fact that $f(\lambda)$ cannot be made negative by the first order terms renders this claim irrelevant, it is necessary to evaluate the contribution of the second order terms when the derivation is corrected by  imposing $\delta M=M\zeta$. To second order in $\lambda$ we have
\begin{equation}
f(\lambda)\sim O(\epsilon^2) - O(\lambda \epsilon \zeta) + O(\lambda^2\zeta^{2})
\label{manifest3}
\end{equation}
It is manifest in (\ref{manifest3}) that the contribution of the second order perturbations vanishes in (\ref{swmain}), as it becomes fourth order when multiplied by the square of the small parameter $\lambda$. (Note that the leading term in (\ref{manifest3}) --which is zeroth order in $\lambda$-- is actually second order in $\epsilon$.) In that respect it is not possible to incorporate the effect of the second order perturbations into the analysis using the SW method. Moreover, when the analysis is corrected by imposing the fact that $\delta M$ is a first order quantity,  even the first order perturbations $(\delta M, \delta J, \delta Q)$ do not contribute to $f(\lambda)$ as one can observe in (\ref{manifest2}) and (\ref{manifest3}).

In the previous works by Hubeny, Jacobson-Sotiriou, and D\"{u}zta\c{s}-Semiz, $\delta_{\rm{fin}}$ defined in (\ref{second}) is made negative for nearly extremal black holes which corresponds to making $f(\lambda)$ negative in the derivation of SW \cite{hu,js,overspin}. The numerical value of $\delta_{\rm{fin}}$  turns out to be of the order $-M^2\epsilon^2$ which suggests that the destruction of the event horizon can be fixed by the second order corrections due to the backreaction effects. Later, these corrections were indeed achieved by employing backreaction effects \cite{backhu,backjs,kerrmog}. The SW method does not reproduce any of these results when one imposes the fact that $\delta M$ is a first order quantity for test bodies and fields. One observes that the terms first order in $\lambda$, contribute to $f(\lambda)$ to third order so they cannot make $f(\lambda)$ negative. The terms that are second order in $\lambda$, cannot fix anything since their contribution is of the fourth order. Therefore the SW method cannot be used to evaluate the effect of second order perturbations. Backreactions have to be explicitly calculated.

\section{Re-visiting the over-spinning problem}
\label{backreaction}
In this section we re-visit the over-spinning problem for Kerr-Newman black holes and explicitly calculate the backreaction effects, which supervenes on the argument that they cannot be calculated using the SW method. We start by attempting to overspin a nearly extremal Kerr-Newman  black hole parametrised as (\ref{param}), by a neutral, scalar test field with frequency $\omega$ and azimuthal wave number $m$. We adapt the parametrization (\ref{testfield}) so that  the field carries energy $\delta M=M\zeta$, in accord with the test field approximation.  At the end of the interaction, the final parameters of the space-time satisfy:
\begin{eqnarray}
M_{\rm{fin}}&=&M+M\zeta \nonumber \\
J_{\rm{fin}}&=&J+\frac{m}{\omega} M\zeta \nonumber \\
Q_{\rm{fin}}&=&Q
\end{eqnarray} 
where $M,J,Q$ are the initial parameters which satisfy (\ref{param}). Now, we demand that the black hole is overspun at the end of the interaction; i.e. $\delta_{\rm{fin}}<0$
\begin{equation}
\delta_{\rm{fin}}=(M+M\zeta)^2-Q^2-\frac{(J+\frac{m}{\omega} M\zeta)^2}{(M+M\zeta)^2}<0
\label{deltafinnex}
\end{equation}
We can substitute  $Q^2=M^2-J^2/M^2-M^2\epsilon^2$ by using (\ref{param}). Re-arranging (\ref{deltafinnex}),
we get
\begin{equation}
M^2 \left(\zeta^2 +2\zeta + \epsilon^2 +\frac{J^2}{M^4} \right)<\frac{J+\frac{m}{\omega} M\zeta)^2}{M^2(1+\zeta)^2}
\label{deltafinnex2}
\end{equation}
We define the dimensionless parameter 
\begin{equation}
\alpha \equiv J/M^2
\label{definealpha}
\end{equation}
Note that for a nearly extremal Kerr-Newman black hole  parametrised as (\ref{param}), the sum
$J^2/M^2 + Q^2$ has a fixed value which is equal to $M^2(1-\epsilon^2)$ for a fixed mass $M$. However nearly extremal black holes satisfying (\ref{param}) may have different values of angular momentum and charge keeping the sum  $J^2/M^2 + Q^2$ fixed. We use the dimensionless parameter $\alpha$ to identify different Kerr-Newman black holes --with a fixed mass $M$-- that all satisfy (\ref{param}). Also note that, substituting $\alpha \equiv J/M^2$ the parametrisation (\ref{param}) can be re-written in terms of the dimensionless variable $\alpha$.
\begin{equation}
1-\alpha^2-\frac{Q^2}{M^2}=\epsilon^2
\label{paramalpha}
\end{equation}
We proceed by taking the square root of both sides of (\ref{deltafinnex2}). The condition $\delta_{\rm{fin}}<0$ reduces to
\begin{equation}
\omega<\omega_{\rm{max}}=\frac{m\zeta}{M\left[ (1+\zeta)\sqrt{\zeta^2+2\zeta+\epsilon^2+\alpha^2}-\alpha\right]}
\label{deltafinnex3}
\end{equation}
We have considered the interaction of a nearly extremal black hole parametrised as (\ref{param}) with a test field carrying energy $\delta M=M\zeta$ and angular momentum $\delta J=(m/\omega)\delta M$. Note that $\delta J$ is inversely proportional to the frequency $\omega$. The equation (\ref{deltafinnex3}) implies that a test field with a frequency $\omega<\omega_{\rm{max}}$, will contribute to the angular momentum parameter of the black hole with a magnitude sufficiently larger than its contribution to the energy parameter so that the final parameters of the black hole describe a naked singularity with $\delta_{\rm{fin}}<0$. In that case we could conclude that the nearly extremal Kerr-Newman black hole is overspun into a naked singularity. However, we should also demand that the test field is absorbed by the black hole, i.e. $\omega$ is larger than the limiting frequency for superradiance, which we denote by $\omega_{\rm{sl}}$. For a nearly extremal black hole parametrised as (\ref{param}), which is perturbed by a neutral test field $(\delta Q=0)$, the superradiance limit is given by
\begin{equation}
\omega_{\rm{sl}}=\frac{ma}{r_+^2 + a^2}=\frac{m}{M\left[\frac{(1+\epsilon)^2}{\alpha}+\alpha\right]}
\label{wslnext}
\end{equation}
Kerr-Newman black holes with different values of $\alpha$ defined in (\ref{definealpha}), have different superradiance limits. For lower values of $\alpha$ which describe black holes with  relatively low angular momentum, the superradiance limit will also be low. In that case the absorption of the modes with relatively low frequencies will be allowed.  The test fields with frequency in the range $\omega_{\rm{sl}}<\omega<\omega_{\rm{max}}$ simultaneously satisfy the two conditions that the field is absorbed by the black hole and it contributes to the angular momentum parameter with a sufficiently large magnitude to overspin the black hole into a naked singularity. We can conclude that the test fields with energy $\delta M=M \zeta$ and frequency in the range $\omega_{\rm{sl}}<\omega<\omega_{\rm{max}}$   can be used to overspin a nearly extremal Kerr-Newman black hole into a naked singularity, provided that $\omega_{\rm{sl}}<\omega_{\rm{max}}$ . Comparing (\ref{deltafinnex3}) and (\ref{wslnext}) one observes that the upper limit for the frequency of the incident field $\omega_{\rm{max}}$ derived in (\ref{deltafinnex3}), is larger than the superradiance limit $\omega_{\rm{sl}}$ for any value of $\alpha$, in the relevant range $(0,1)$. It turns out that every nearly extremal Kerr-Newman black hole satisfying (\ref{param}) can be overspun by neutral test fields, regardless of the specific value of $\alpha$ defined in (\ref{definealpha}). The validity of this conclusion is limited to the case, where one ignores the backreaction effects. 
\subsection{Backreactions for nearly extremal black holes}
\label{backreaction_nex}
In this derivation we have not ignored the contribution of the second order terms $(\delta M)^2$ and $(\delta J)^2$. Therefore  we have to employ the backreaction effects to test whether the destruction of the event horizon can be fixed. Since, the Sorce-Wald method is invalid we have to explicitly determine and calculate the backreaction effects. Backreaction effects will bring second order corrections to $\delta_{\rm{fin}}$ which can in principle restore the event horizon. The most legitimate type of backreaction effect for an overspinning problem is the induced increase in the angular velocity of the event horizon before the absorption of the test body/field occurs, which was suggested by Will \cite{will}.  The induced increase in the angular velocity of the event horizon leads to an increase in the superradiance limit. This implies that the absorption of the challenging modes with relatively low frequencies can be prevented. In a recent paper we have employed this backreaction effect for the overspinning problem of Kerr-MOG black holes \cite{kerrmog}. 

We envisage a test field with angular momentum $\delta J$ incident on a black hole with mass $M$. According to the estimate in \cite{will}, the angular velocity of the event horizon increases by an amount
\begin{equation}
\Delta \omega =\frac{\delta J}{4M^3}
\label{deltaomegamain}
\end{equation}
The increase in the angular velocity of the event horizon results in an increase in the superradiance limit, which will be modified as
\begin{equation}
\omega_{\rm{sl}}'=\omega_{\rm{sl}}+\Delta \omega
\label{wslmodified}
\end{equation}
In (\ref{deltafinnex3}) we have derived the maximum value for the frequency of a test field that could overspin a nearly extremal Kerr-newman black hole parametrised as (\ref{param}). We noted that the fields with frequency in the range $\omega_{\rm{sl}}<\omega<\omega_{\rm{max}}$ can lead to overspinning. However as the test field is incident on the black hole the angular velocity of the event horizon will increase, which will lead to a modification in the superradiance limit given derived in (\ref{wslmodified}). If the modified value of the superradiance limit exceeds the frequency of the incoming field, the absorption of the test field is prevented and the event horizon cannot be destroyed. The test field will be scattered back to infinity with a larger magnitude. Note that $\delta J$ and $\Delta \omega$ given in (\ref{deltaomegamain}) are inversely proportional to the frequency $\omega$. For that reason,   if the modified value of the superradiance limit exceeds the incoming frequency for $\omega \simeq \omega_{\rm{max}}$, it will exceed the incoming frequency even further for smaller values in the range $\omega_{\rm{sl}}<\omega<\omega_{\rm{max}}$, as we have argued in \cite{kerrmog}. Therefore it is critical to calculate $\Delta \omega$ for the frequencies arbitrarily close to $\omega_{\rm{max}}$.  

To calculate backreactions, we first envisage a test field with  frequency arbitrarily close to but slightly less than $\omega_{\rm{max}}$, incident on a nearly extremal Kerr-Newman black hole parametrised as (\ref{param}). The test field carries energy $\delta M=M \zeta$ and angular momentum $\delta J=(m/\omega) \delta M$, where $\omega \lesssim \omega_{\rm{max}}$. According to the derivation in the previous section, this test field will lead to the overspinning of the nearly extremal Kerr-Newman black hole. Now we incorporate the backreaction effects due to the induced increase in the angular velocity of the event horizon. We check if the modified value of the superradiance limit derived in (\ref{wslmodified}), exceeds the frequency of the incoming field.
For simplicity we let $\epsilon \simeq \zeta$ and substitute $\omega =\omega_{\rm{max}}$  in (\ref{deltaomegamain}) to calculate $\Delta \omega$.
\begin{equation}
\Delta \omega =\frac{\left(\frac{m}{\omega}\right)M \epsilon}{4M^3}=\frac{\left[(1+\epsilon)\sqrt{2\epsilon^2 + 2\epsilon +\alpha^2}-\alpha\right]}{4M}
\label{deltaomega}
\end{equation}
For $\omega \simeq \omega_{\rm{max}}$, the increase in the superradiance limit is given by $\Delta \omega$ in (\ref{deltaomega}), which leads to the modified value of the superradiance limit denoted by $\omega_{\rm{sl}}'$ derived in (\ref{wslmodified}). Then, we need to compare $\omega_{\rm{sl}}'$ and $\omega_{\rm{max}}$. If $\omega_{\rm{sl}}'$ is larger than $\omega_{\rm{max}}$, no net absorption of the test field will occur and overspinning will be prevented.

It turns out that $\omega_{\rm{sl}}'$ defined in (\ref{wslmodified}) is indeed larger than $\omega_{\rm{max}}$ provided that 
\begin{equation}
\alpha \gtrsim 0.50
\label{alphanext}
\end{equation}
Note that the parameter $\alpha \equiv J/M^2$ defined in (\ref{definealpha}) is used to distinguish different nearly extremal Kerr-Newman black holes that all satisfy (\ref{param}). These black holes may have different angular momentum and charge parameters keeping the sum $(J^2/M^2 + Q^2)$ fixed. Without employing bakreaction effects, one derives that every nearly extremal Kerr-Newman black hole can be overspun by test fields, regardless of the specific value of $\alpha$. This includes the Kerr limit $Q \to 0$, $\alpha^2 \to (1-\epsilon^2)$. When one employs backreaction effects due to the induced increase in the superradiance limit, it turns out that there are two there are two possibilities depending on the specific value of $\alpha$. If  $\alpha<0.5$, the modified value of the superradiance limit ($\omega'_{\rm{sl}}$), will still be smaller than $\omega_{\rm{max}}$.  The absorption of the modes in the range $\omega'_{\rm{sl}}<\omega<\omega_{\rm{max}}$ will lead to overspinning. However if $\alpha > 0.5$, the increase in the superradiance limit will be sufficient to prevent the absorption of all challenging modes. Since $\omega'_{\rm{sl}}$ is larger than $\omega_{\rm{max}}$ for this class of nearly extremal Kerr-Newman black holes, all the modes with $\omega<\omega_{\rm{max}}$ that could potentially overspin the black hole will be reflected back to infinity without any net absorption. Thus, the backreaction effects will prevent overspinning. The argument is also valid in the Kerr limit $Q \to 0$. For nearly extremal Kerr black holes, the parameter $\alpha$ has a unique value which is equal to $\sqrt{1-\epsilon^2}$. Manifestly, $\alpha > 0.5$ for nearly extremal Kerr black holes and the overspinning problem is fixed by employing backreaction effects.

It would be appropriate to give a numerical example to elucidate the subject. For that purpose,  let us consider a nearly extremal Kerr-Newman black hole with $\alpha=0.5$ and $\epsilon=0.01$. Initially the parameters of this black hole satisfy
\[
M^2-(J^2/M^2)-Q^2=(0.0001)M^2 \]
or equivalently
\[1-\alpha^2-(Q^2/M^2)=0.0001
\]
For $\alpha=0.5$ the initial parameters of the black hole are given by $J=0.5M^2
$ and $Q^2=0.7499M^2$. Now we perturb this black hole with a test field with energy $\delta M=M\zeta$ and angular momentum $\delta J=(m/\omega) \delta M$. To choose the frequency of the test field we calculate the critical values. We choose $m=1$ which is the mode with the highest probability of absorption. We use equation (\ref{wslnext}) to calculate the superradiance limit.
\[\omega_{\rm{sl}}=0.39366 (1/M)
\]
Letting $\zeta=\epsilon=0.01$, we can find the maximum value for the frequency of the test field that could overspin the black hole, which is analytically derived in (\ref{deltafinnex3})
\[ \omega_{\rm{max}}=0.399908 (1/M) 
\]
If we choose the frequency of the incoming field in the range $\omega_{\rm{sl}}<\omega<\omega_{\rm{max}}$, the test field with energy $\delta M=0.01 M$ will be absorbed by the black hole and it will lead to overspinning. For example let us choose 
\[ \omega=0.395 (1/M)\]
Using $\delta M=0.01M$ and we can calculate $\delta J$
\[
\delta J =(m/\omega) \delta M=0.025316 M^2
\]
The final parameters of the black hole satisfy
\[
\delta_{\rm{fin}}=(M+\delta M)^2-\frac{(J+\delta J)^2}{(M+\delta M)^2}-Q^2=-000319M^2
\]
Note that $Q^2=0.7499M^2$ for $\alpha=0.5$. The fact that $\delta_{\rm{fin}}$ is negative imply that the final parameters of the black hole describe a naked singularity. One can choose any value in the range $\omega_{\rm{sl}}<\omega<\omega_{\rm{max}}$, and verify that $\delta_{\rm{fin}}$ is negative.

Now we employ the backreaction effects due to the induced increase in the angular velocity of the horizon. We have argued that the increase in the angular velocity of the horizon leads to an increase in the superradiance limit, which will prevent the absorption of the challenging modes for a class of nearly extremal Kerr-Newman black holes with $\alpha \gtrsim 0.5$. The increase in the superradiance limit $(\Delta \omega)$ is analytically derived in (\ref{deltaomega}). Note that $(\Delta \omega)$ is inversely proportional to the frequency of the incoming field so the minimum increase occurs for $\omega \simeq \omega_{\rm{max}}$ considering the challenging modes. Using (\ref{deltaomega}) we calculate the minimum increase in the superradiance limit for this black hole
\[
\Delta \omega =0.00625 (1/M) 
\]
With this increase the superradiance limit is modified as
\[
\omega_{\rm{sl}}'=0.39991 (1/M)
\]
which implies that the modes with $\omega <0.39991 (1/M)$ will not be absorbed by the black hole. Since this value is larger than $\omega_{\rm{max}}$, none of the challenging modes will be absorbed by the black hole. Thus, the increase in the superradiance limit prevents overspinning as no net absorption of the challenging modes occur.

However for smaller values of $\alpha$, $\omega_{\rm{sl}}'$ will still be less than $\omega_{\rm{max}}$. Then, the frequencies in the range $\omega_{\rm{sl}}' < \omega < \omega_{\rm{max}}$ can be used to overspin the nearly extremal black hole. The increase in the angular velocity of the event horizon does not bring an ultimate solution to the overspinning problem for Kerr-Newman black holes. However, backreactions is an open problem. Various different forms of backreactions can be considered which can possibly fix the overspinning problem.
\subsection{Backreactions for extremal black holes}
\label{backreaction_ex}
By definition, the initial parameters of extremal Kerr-Newman black holes satisfy:
\begin{equation}
M^2-Q^2-(J^2)/(M^2)=0
\label{mainext}
\end{equation}
or equivalently
\begin{equation}
1-\alpha^2-(Q^2/M^2)=0
\label{mainextalpha}
\end{equation}
If an extremal Kerr-Newman black hole is perturbed by a neutral test field, the limiting frequency for superradiance is
\begin{equation}
\omega_{\rm{sl-ex}}=\frac{ma}{r_+^2 + a^2}=\frac{m}{M\left( \frac{1}{\alpha}+ \alpha\right)}
\label{superradex}
\end{equation}
In a recent paper we have shown that $\delta_{\rm{fin}}$ becomes negative for an extremal Kerr-Newman black hole if the frequency of the test field is less than the maximum value \cite{generic}
\begin{equation}
\omega<\omega_{\rm{max-ex}}=\frac{m\zeta}{M\left[ (1+\zeta)\sqrt{\zeta^2+2\zeta+\alpha^2}-\alpha\right]}
\label{omegamaxext}
\end{equation}
which is the $\epsilon \to 0$ limit of the value derived for nearly extremal black holes in (\ref{deltafinnex3}).  Again the two conditions should be satisfied simultaneously for overspinning to occur. The test field should be absorbed by the black hole and its contribution to the angular momentum parameter should be larger that the contribution to the mass parameter. In \cite{generic} we have shown that $\omega_{\rm{sl-ex}}$ is less than $\omega_{\rm{max-ex}}$ for a class of extremal Kerr-Newman black holes which satisfy
\begin{equation}
\alpha^2 \equiv \frac{J^2}{M^4}<\frac{1}{3} \Rightarrow \alpha \lesssim 0.577
\label{extrem}
\end{equation}
This implies that the test fields with frequency in the range  $\omega_{\rm{sl-ex}}< \omega<\omega_{\rm{max-ex}}$ can be used to overspin the extremal Kerr-Newman black holes which satisfy (\ref{extrem}). The derivation is incomplete as it ignores the contribution of the backreaction effects. As in the case of nearly extremal black holes we calculate the induced increase in the angular velocity of the event horizon before the absorption of the test field.
\begin{equation}
\Delta \omega =\frac{\delta J}{4M^3}=\frac{(m/\omega)M\zeta}{4M^3}
\label{deltaomegaex}
\end{equation}
By substituting $\omega=\omega_{\rm{max-ex}}$ in (\ref{deltaomegaex}) we derive that 
\begin{equation}
\Delta \omega =\frac{1}{4M}\left[ (1+\zeta)\sqrt{\zeta^2+2\zeta+\alpha^2}-\alpha\right]
\label{deltaomegaext}
\end{equation}
We should add the induced increase in the angular momentum of the extremal Kerr-Newman black hole derived in (\ref{deltaomegaext}) to the superradiance limit (\ref{superradex}). This gives us the modified value of the superradiance limit. As we have argued in the previous section, we demand that the modified value of the superradiance limit is larger than $\omega_{\rm{max-ex}}$ derived in (\ref{omegamaxext}), so that the absorption of all the challenging modes with $\omega <\omega_{\rm{max-ex}}$ is prevented. For that purpose we let $\zeta=0.01$, and demand that
\begin{equation}
\omega_{\rm{sl-ex}}+\Delta \omega > \omega_{\rm{max-ex}}
\label{demandext}
\end{equation}
One derives that (\ref{demandext}) is satisfied, i.e. the modified value of superradiance will exceed the maximum value of the frequency of the test field $\omega_{\rm{max-ex}}$, provided that 
\begin{equation}
\alpha \gtrsim 0.31
\label{extresult}
\end{equation}
The dimensionless parameter $\alpha$ was introduced to distinguish extremal Kerr-Newman black holes with different angular momentum and charge parameters that all satisfy (\ref{mainext}) and its dimensionless equivalent (\ref{mainextalpha}). In a recent paper we derived that extremal Kerr-Newman black holes which satisfy $\alpha \lesssim 0.57 $ can be overspun by test fields \cite{generic}. The employment of backreaction effects bring a further restriction to the class of extremal Kerr-Newman black holes that can be overspun by test fields. The result (\ref{extresult}) implies that the absorption of all the challenging test fields will be prevented due to the induced increase in the superradiance limit, provided that $\alpha \equiv (J)/(M^2)\gtrsim 0.31$. Let us elucidate the subject with a numerical example. For that purpose let us consider an extremal Kerr-Newman black hole with $\alpha=0.32$. The initial parameters of this black hole satisfy $J=0.32 M^2$ and $Q^2=0.8976 M^2$ so that the black hole is extremal. Our first claim is that the black hole can be overspun by test fields if one ignores the backreaction effects. To verify this claim let us first calculate the critical values   $\omega_{\rm{sl-ex}}$ and $\omega_{\rm{max-ex}}$ which were analytically derived in (\ref{superradex}) and (\ref{omegamaxext}).
\begin{eqnarray}
& &\omega_{\rm{max-ex}}=0.298507 (m/M) \nonumber \\
& & \omega_{\rm{sl-ex}}=0.290276 (m/M)
\end{eqnarray}
We claim that if we choose a test field with frequency in the range $\omega_{\rm{sl-ex}}<\omega<\omega_{\rm{max-ex}}$ and energy $\delta M=M\zeta$, it will be absorbed by the extremal black hole and overspin it into a naked singularity. Let us choose a test field with
\[
\omega=0.295 (m/M); \quad  \delta M=M\zeta=0.01 M 
\]
\[ \delta J=\frac{m}{\omega}\delta M=0.033898M^2
\]
By definition $\delta_{\rm{in}}=0$ for an extremal black hole. Let us calculate $\delta_{\rm{fin}}$.
\[
\delta_{\rm{fin}}=(M+\delta M)^2-\frac{(J+\delta J)^2}{(M+\delta M)^2}-Q^2=-0.000276M^2
\]
The negative value of $\delta_{\rm{fin}}$ implies that the black hole is overspun. However the fact that $\delta_{\rm{fin}} \sim \zeta^2$ indicates that the overspinning can be fixed by backreaction effects.

Our second claim is that the overspinning of this extremal Kerr-Newman black hole should be fixed by the induced increase in the superradiance limit since $\alpha>0.31$.  Using the analytical result (\ref{deltaomegaext}), we calculate that, the limiting frequency for superradiance to occur will increase by an amount
\begin{equation}
\Delta \omega=0.008375 (1/M)
\end{equation}
Then, for $m=1$ the modified value of the superradiance limit will be
\begin{equation}
\omega'_{\rm{sl-ex}}=\omega_{\rm{sl-ex}}+\Delta \omega=0.298651 (1/M)
\end{equation}
Since the modified value of the superradiant limit exceeds $\omega_{\rm{max-ex}}$, the test fields which could overspin the black hole will not be absorbed by the black hole. In particular the test field that we have chosen for our numerical example will not be absorbed, since $\omega=0.295 (1/M)<0.298651 (1/M)$. However for smaller values of $\alpha$ the modified value of the superradiance limit will still be less than $\omega_{\rm{max-ex}}$. The induced increase in the angular velocity of the event horizon brings further restrictions to the class of extremal Kerr-Newman black holes which can be overspun by test fields, though it does not completely fix the problem. As in the case of nearly extremal black holes, we note that different type of backreaction effects can be employed to fix the problem.

\section{Absorption Probabilities in Wald type problems}
\label{prob}
As we stated in the introduction, the interaction of black holes with test fields is actually a scattering problem. The test fields are partially absorbed by the black hole, and partially reflected back to infinity. For classical fields, t
he transmission and reflection coefficients represent the ratios of energies that are respectively, absorbed by the black hole and scattered back to infinity. The conservation of energy implies that the sum of the coefficients should be unity.
\begin{equation}
\frac{\Phi_{\rm{reflected}}}{\Phi_{\rm{incident}}} + \frac{\Phi_{\rm{transmitted}}}{\Phi_{\rm{incident}}}=1
\label{flux}
\end{equation} 
Conventionally the relative flux $(\Phi_{\rm{transmitted}})/(\Phi_{\rm{incident}})$ is interpreted as the absorption probability of the incident field. This notion would be improper for the cases of superradiant scattering. If the frequency of the incoming field is lower than the superradiance limit $(\omega<\omega_{\rm{sl}})$, 
the wave carries energy out of the black hole and the absorption probability will be negative. The conservation of energy described by the equation (\ref{flux}) continues to hold. Though the notion of a negative probability can be improper, we shall continue adapt the conventional term ``absorption probability'' for the relative flux $(\Phi_{\rm{transmitted}})/(\Phi_{\rm{incident}})$, throughout this paper.

In all the previous Wald  type problems to test the stability of event horizons, the effect of the absorption probability of the test fields has been ignored. The works of this author are not  exceptions to this general attitude. Ignoring the absorption probability corresponds to assuming that the probability is of the order of unity if the field is absorbed by the black hole.  However the test fields are partially absorbed by the black hole and partially reflected back to infinity. Only the transmitted part of the test field contributes to the mass, angular momentum, and charge parameters of the black hole. In this sense, the magnitude of the contribution is directly proportional to the absorption probability $(\Phi_{\rm{transmitted}})/(\Phi_{\rm{incident}})$. For the challenging modes, the absorption probability approaches zero as the frequency becomes close to the superradiance limit. This fundamentally alters the course of the analysis for the interaction of test fields with extremal and nearly extremal black holes.  

In this work we incorporate the effect of absorption probabilities into Wald type problems. For that purpose, we modify the contributions of a test field to the mass, angular momentum, and charge parameters of the black hole taking the absorption probabilities into consideration. If a test field carries energy $M \zeta$ and the absorption probability of the field is $\Gamma$, the energy absorbed by the black hole will be 
\begin{equation}
E_{\rm{abs}}=\Gamma  (M \zeta)
\label{deltaeabs}
\end{equation}
while the energy reflected back to infinity is 
\begin{equation}
E_{\rm{ref}}=(1-\Gamma)(M \zeta)
\label{deltaeref}
\end{equation}
The expressions (\ref{deltaeabs}) and (\ref{deltaeref}) are direct consequences of the fact that only the transmitted part of the test field contribute to the parameters of the black hole. As we have mentioned above, in all the previous problems, the field is assumed to be entirely absorbed by the black hole (the absorption probability of the field is $\Gamma$ is assumed to be of the order of unity) so that $\delta M=M\zeta $.  After a sufficiently long time, the mass and angular momentum parameters of the black hole will be modified as:
\begin{eqnarray}
M_{\rm{final}}&=& M +E_{\rm{abs}}=M+ \Gamma (M\zeta) \nonumber \\
J_{\rm{final}}&=& J + J_{\rm{abs}}=J+\frac{m}{\omega} \Gamma ( M \zeta)
\label{modified}
\end{eqnarray}
where $E_{\rm{abs}}$ and $J_{\rm{abs}}$ are the energy and the angular momentum absorbed by the black hole. The absorption probability $\Gamma$ can be positive, negative or zero. The absorption probability $\Gamma$ appearing in (\ref{modified}) should not be confused with the small parameter $\lambda$ introduced by Sorce and Wald. $\Gamma$ is not necessarily a small parameter. It can be of the order of unity for test fields with frequency $\omega \gg \omega_{\rm{sl}}$. It is identically zero for the optimal perturbations with  $\omega = \omega_{\rm{sl}}$, and it is negative for the test fields in the superradiant range  $\omega < \omega_{\rm{sl}}$

The absorption probabilities $\Gamma_{s\omega lm}$ for the wave modes with spin $s$, frequency $\omega$, spheroidal harmonic $l$ and azimuthal wave number $m$, were first calculated by Page \cite{page}. The absorption probability of the wave depends on the parameters of the black hole such as the mass $M$, the charge $Q$, the angular momentum $J$, the area of the event horizon $A$, the surface gravity $\kappa$, the angular velocity of the event horizon $\Omega$ and the electrostatic potential of the event horizon $\Phi$. The parameters of the black hole are not independent. In particular The area and the surface gravity satisfy
\[
\kappa A=4\pi (r_+ - M)
\]
where $(r_+ - M)=\sqrt{M^2-a^2-Q^2}$ for a Kerr-Newman black hole. In this work we are interested in the absorption probability of scalar fields ($s=0$) incident on Kerr-Newman black holes. The modes with $m=0$ do not contribute to angular momentum, therefore we should consider the modes with $m \geq 1$. For $l=1$ Page's results imply that
\begin{equation}
\Gamma_{0\omega 1 m}=\frac{1}{9}\frac{A}{\pi}[M^2-(m^2-1)a^2-Q^2](\omega-m\Omega)\omega^3
\label{prob1}
\end{equation}
The factor $(\omega-m\Omega)$ implies that  the absorption probability  will be negative if the incident field is in a superradiant mode $(\omega<\omega_{\rm{sl}})$. In that case the mass of the black hole will decrease. However, the angular momentum will decrease by a much larger magnitude and $\delta_{\rm{fin}}$ will be positive. Therefore the modes in the superradiant range $(\omega<\omega_{\rm{sl}})$ do not lead to overspinning.

\subsection{Optimal perturbations}
\label{optimal}
Optimal perturbations satisfy the inequality (\ref{needham}) at the lower limit  so that
\begin{equation}
\delta M=\Omega \delta J + \phi \delta Q
\label{optimal1}
\end{equation}
This constitutes the lower limit to allow the absorption of a test body or field.  For neutral  test fields with energy $\delta M$ and angular momentum $\delta J=(m/\omega)\delta M$, (\ref{optimal1}) implies that the frequency of the test field is equal to the superradiance limit $(\omega=\omega_{\rm{sl}}=m\Omega)$ for the optimal perturbations. Since the test bodies and fields with lower energies than the optimal perturbations are not absorbed by the black holes, they need not be considered for the overspinning and overcharging problems. The optimal perturbations carry the lowest possible energy relative to their angular momentum and charge. Thus,  when one ignores absorption probabilities, these modes appear to be more likely to lead to overspinning or overcharging than any other mode that is absorbed by the black hole.  

However when one incorporates the absorption probabilities into the problem, the course of the analysis is fundamentally altered. It is  manifest in (\ref{prob1}) that the absorption probability is zero for the optimal perturbations with $\omega = \omega_{\rm{sl}}=m\Omega$. In that case the field is entirely reflected back to infinity, with the same amplitude. No net absorption of the field occurs. The final parameters of the spacetime given by (\ref{modified}) are identically equal to the initial parameters. 
\begin{eqnarray}
&&M_{\rm{final}}= M + \Gamma ( M \zeta)=M \nonumber \\
&&J_{\rm{final}}= J + \frac{m}{\omega} \Gamma ( M \zeta)=J
\end{eqnarray}
Therefore the optimal perturbations do not challenge the stability of the event horizon. Whether we start with an extremal or a nearly extremal black hole, we end up with the same black hole surrounded by an event horizon. The parameters of the black hole remain invariant after the interaction. The black hole maintains its initial state. When absorption probability is taken into account, the optimal perturbations become irrelevant for the overspinning and overcharging problems. 

\subsection{Nearly extremal black holes and challenging modes}
\label{prob_nex}
The modes that could potentially overspin black holes have frequencies close to the superradiance limit. The absorption probabilities of these modes are approach zero as one approaches the superradiance limit. By definition, a test field carries a small amount of energy and angular momentum and only a small fraction of its energy and angular momentum will be absorbed by the black hole if its frequency is close to the superradiance limit. Therefore it seems to be very difficult for a test field to drive a nearly extremal black hole to extremality and beyond, when the absorption probability is taken into account. To evaluate this quantitatively,  let us consider a scalar field with frequency 

\begin{equation}
\omega=m\Omega(1+\xi)
\label{definexi}
\end{equation}
where the small parameter $\xi \ll 1$ assures that the frequency of the incoming field is close to the superradiance limit $\omega_{\rm{sl}}=m\Omega$. The scalar field is incident on a nearly extremal Kerr-Newman black hole parametrised as (\ref{param}), where $\epsilon \ll 1$ determines the closeness of the black hole to extremality. The highest absorption probability occurs for $m=1$. Substituting $\omega=\Omega(1+\xi)$ in (\ref{prob1})
\begin{eqnarray}
\Gamma_{0\omega 1 1}&=&\frac{8}{9} M^2(1+\epsilon)[M^2-Q^2](\Omega \xi)[\Omega(1+\xi)]^3 \nonumber \\
&\sim & \xi + O(\epsilon \xi)
\label{prob2}
\end{eqnarray}
where we have used $A=8\pi M^2 (1+\epsilon)$ for a nearly extremal black hole. The leading term in the absorption probability of a challenging mode is of the order of $\xi$. Now, we can re-evaluate the overspinning problem taking the absorption probability into consideration. We start with a Kerr-Newman black hole satisfying
\[
M^2-Q^2-\frac{J^2}{M^2}=M^2 \epsilon^2 
\]
which is perturbed by a test field with
\begin{eqnarray}
&&m=1; \quad \omega= m\Omega(1+\xi) \nonumber \\
&& \delta M=M\zeta ; \quad \delta J=\frac{1}{\omega}\delta M; \quad \delta Q=0 \nonumber \\
&& \Gamma \sim \xi
\end{eqnarray}
We should note that we choose the energy, frequency, and the azimuthal wave number of the test field to challenge the stability of the event horizon. The absorption probability of this test field is not a choice; it follows from (\ref{prob1}). After the interaction of the test field with the Kerr-Newman black hole, the final parameters of the space-time will take the form
\begin{eqnarray}
M_{\rm{final}}&=& M + M\zeta \xi \nonumber \\
J_{\rm{final}}&=& J+\frac{1}{\omega} M\zeta \xi  \nonumber \\
Q_{\rm{final}}&=&Q
\label{finalparamprob}
\end{eqnarray}
We demand that the final parameters of the space-time describe a naked singularity.
\begin{equation}
(M+M \zeta \xi )^2 - Q^2 -\frac{\left( J+\frac{1}{\omega}M \zeta \xi  \right)^2}{(M+M \zeta \xi )^2} < 0
\label{probcondi1}
\end{equation}
As in the previous sections, we eliminate $Q$ from (\ref{probcondi1}), and the define the dimensionless variable $\alpha=J/M^2$. After some algebra one derives that the condition (\ref{probcondi1}) is equivalent to
\begin{equation}
\omega < \omega_{\rm{max}}=\frac{ \zeta \xi }{M \left[(1+  \zeta \xi )\sqrt{2 \zeta \xi  + \epsilon^2 + \alpha^2}-\alpha \right]}
\label{omegamaxprob}
\end{equation}
We have assumed that the frequency of the incoming field is slightly larger than the superradiance limit by imposing $\Gamma=\xi$. Remember that the superradiance limit for a neutral test field is
\[ \omega_{\rm{sl}}=\frac{m}{M\left[\frac{(1+\epsilon)^2}{\alpha}+\alpha\right]}
\]
The maximum value of the frequency of the incident field derived in (\ref{omegamaxprob}), has to be larger than the superradiance limit. Letting $\epsilon=\zeta= \xi =0.01$, one derives that this will only be possible if
\begin{equation}
\alpha \lesssim 0.01041
\label{probcondi2}
\end{equation}
We started with a nearly extremal Kerr-Newman black hole parametrised as (\ref{param}). We perturbed this black hole with a test field with energy $\delta M=M\zeta$. For overspinning to occur the frequency of the test should be as small as possible since the contribution to the angular momentum is inversely proportional to the frequency. The test field should also be absorbed by the black hole which entails that the frequency should be larger than the superradiance limit. Therefore we the frequency should be slightly larger than the superradiance limit  ($\omega=m\Omega(1+\xi)$)  for the test field. Using (\ref{prob1}) which follows from the seminal results by Page \cite{page}, one can show that the absorption probability for this field is of the order of $\xi$. The final parameters of the black hole are given by (\ref{modified}), which implies that only the fraction of the test field that is absorbed by the black hole contributes to the mass, angular momentum, (and charge) parameters. We demand that the final parameters of the black hole describe a naked singularity; i.e. $\delta_{\rm{fin}}<0$. We derive that this demand is satisfied if the frequency of the field is smaller than the maximum value derived in (\ref{omegamaxprob}). This value is larger than the superradiance limit for a class of Kerr-Newman black holes with $\alpha=J/M^2\lesssim 0.01041$. To clarify the reader, we note that for a nearly extremal Kerr-Newman black hole with mass $M=1$ and $\alpha=0.01$, the angular momentum and charge parameters satisfy $J^2=0.0001$ and $Q^2=0.9998$ with $\epsilon=0.01$. (See equation (\ref{paramalpha}). ) For this class of nearly extremal Kerr-Newman black holes there exist modes with positive absorption probability, that could increase the angular momentum parameter beyond the extremal limit. This stems from the fact that the modes with very low frequencies can have positive absorption probabilities for such low values of $\alpha$. For these modes, the contribution to angular momentum will be very large as it is inversely proportional to the frequency. 
However, the induced increase in the angular velocity of the event horizon will also be very large for these fields. The modified value of the superradiance limit, which was analytically derived in (\ref{wslmodified}), will considerably exceed the frequency of the test field and its absorption will be prevented.

Let us clarify this argument with a quantitative examle. For that purpose we  consider a nearly extremal Kerr-Newman black hole with $\alpha \equiv J/M^2=0.01$. For this black hole, we can use (\ref{omegamaxprob}) and the general expression for the superradiance limit to find that
\begin{eqnarray}
&&\omega_{\rm{max}}=0.009998 \frac{1}{M} \nonumber \\
&&\omega_{\rm{sl}}=0.009802 \frac{1}{M}
\end{eqnarray}
Let us consider a test field with 
\begin{eqnarray}
&&m=1; \quad \omega=0.0099 \frac{1}{M}\sim m\Omega(1+\xi) \nonumber \\
&& \delta M=M \zeta=0.01M \nonumber \\
&& \delta J= \frac{m}{\omega}\delta M=1.0101 M^2 \nonumber \\
&& \Gamma \sim \xi
\end{eqnarray}
Without considering backreaction effects, one derives that the final parameters of the black hole given in (\ref{finalparamprob}) describe a naked singularity rather than a black hole. However, one can notice that $\delta J$ is too large for this field; in particular it is even larger than $M^2$. Let us calculate the modified value of the limiting frequency due to the induced increase in the angular momentum of the horizon, for this mode. Using (\ref{deltaomegamain}) and (\ref{wslmodified}), 
\begin{eqnarray}
&& \Delta \omega =\frac{\delta J}{4M^3}=0.2525 \left( \frac{1}{M} \right) \nonumber \\
&& \omega_{\rm{sl}}'=\omega_{\rm{sl}}+ \Delta \omega =0.262302 \left( \frac{1}{M} \right) 
\end{eqnarray}
The modified value of the superradiance limit is far larger than the frequency of the incoming field, which assures that the test field will not be absorbed by the black hole. Therefore the backreaction effects fix the overspinning problem for nearly extremal Kerr-Newman black holes with $\alpha \equiv J/M^2 \lesssim 0.01041$. Moreover, one can argue that the challenging modes for this class of Kerr-Newman black holes cannot be treated in the test field approximation, since $\delta J \gtrsim M^2$. In this case these modes can be excluded, and the overspinning problem becomes irrelevant. In either case, the incorporation of the absorption probability brings an ultimate solution to the overspinning problem for nearly extremal Kerr-Newman black holes. 
\subsection{Extremal black holes and challenging modes}
\label{prob_ex}
\begin{center}
\begin{figure}
\includegraphics[scale=0.2]{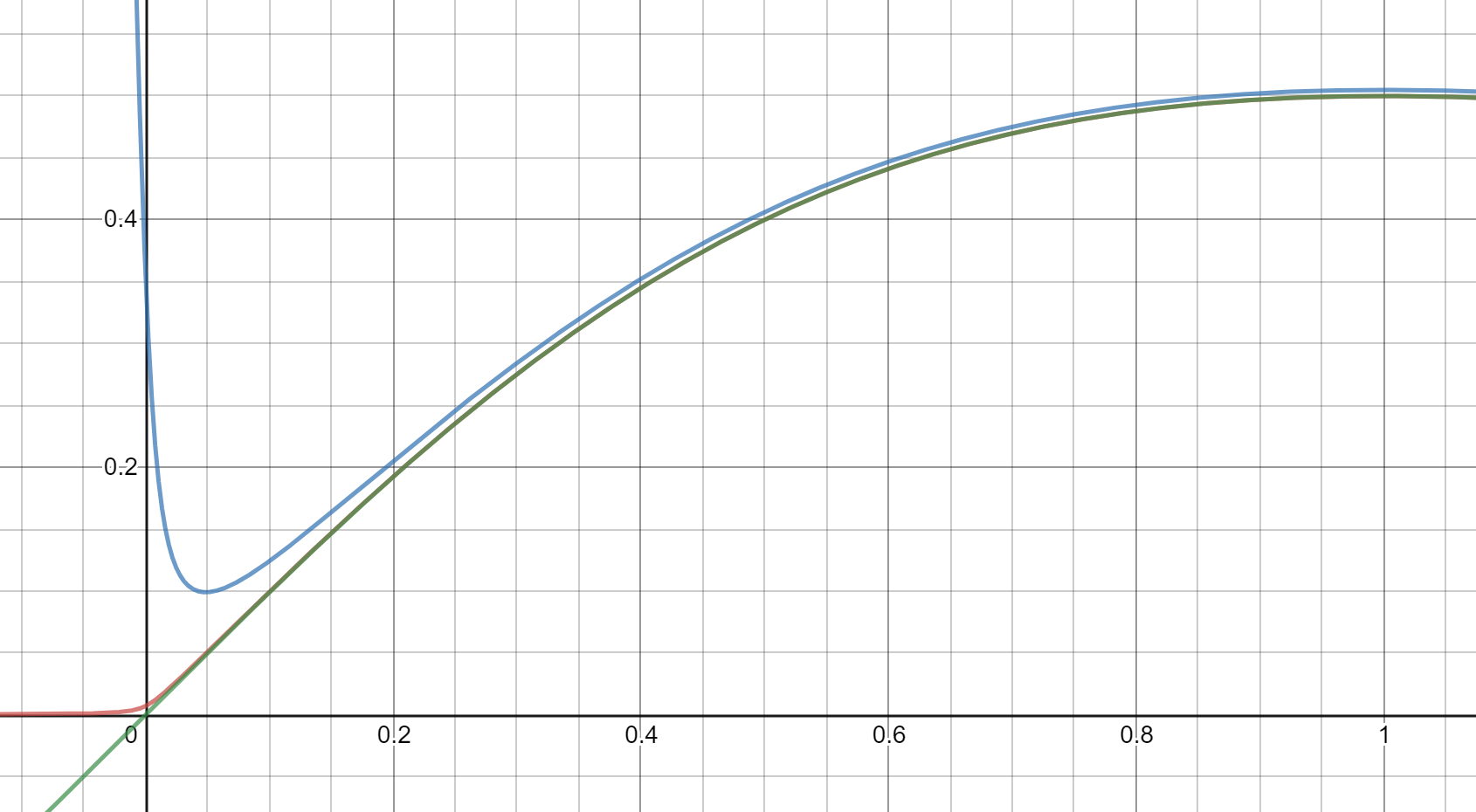}
\caption{The superradiance limit $\omega_{\rm{sl}}$ and the upper limit $\omega_{\rm{max-ex}}$ almost coincide in the range $0<\alpha<1$. The modified value of superradiance $\omega_{\rm{sl}}'$ exceeds the upper limit $\omega_{\rm{max-ex}}$. (Here we let $M=1$)}
\label{figure}
\end{figure}
\end{center}

We mentioned that the absorption probability is very low for the challenging modes. Most of the energy and angular momentum carried by the test field is reflected back to infinity. This leads one to conclude that it should be very difficult for a test field to drive a nearly extremal black hole to extremality and beyond. The situation is different for extremal black holes. A tiny excess amount of angular momentum can lead to overspinning, since $\delta_{\rm{in}}$ is equal to zero. In this section, we calculate the possibility to overspin an extremal Kerr-Newman black hole by neutral test fields by taking the absorption probabilities into consideration. By definition, an extremal black hole satisfies (\ref{mainext}) and its dimensionless equivalent (\ref{mainextalpha}). The dimensionless parameter $\alpha \equiv J/M^2$ allows us to distinguish extremal black holes with different parameters of charge and angular momentum.
As in the case of nearly extremal black holes we attempt to destroy the horizon by sending in a test field with energy $\delta M=M\zeta$, and frequency close to the superradiance limit, such that the absorption probability takes the form $\Gamma \sim \xi$. Proceeding the same way as in the previous section, we choose $m=1$ for the test field which maximizes the absorption probability.  We find that the event horizon will be destroyed if the frequency of the test field satisfies
\begin{equation}
\omega < \omega_{\rm{max-ex}}=\frac{ \zeta \xi}{M \left[(1+  \zeta \xi)\sqrt{2 \zeta \xi  + \alpha^2}-\alpha \right]}
\label{omegamaxprobex}
\end{equation}
which is the $\epsilon \to 0$ limit of the result found in (\ref{omegamaxprob}). Remember that the superradiance limit for a neutral test field incident on an extremal black hole is
\[ \omega_{\rm{sl}}=\frac{1}{M\left[\frac{1}{\alpha}+\alpha\right]}
\]
where $m=1$. Letting $\zeta= \xi=0.01$, one observes that the upper limit $\omega_{\rm{max-ex}}$ derived in (\ref{omegamaxprobex}), and the lower limit $\omega_{\rm{sl}}$ for the range of the frequencies that can lead to overspinning, almost coincide in the range $0<\alpha<1$. The upper limit derived in (\ref{omegamaxprobex}) is slightly larger than the superradiance limit provided that
\begin{equation}
\alpha \lesssim 0.70707
\label{extremalalpha}
\end{equation}
where the dimensionless parameter $\alpha$ defined in (\ref{definealpha}), distinguishes extremal Kerr-Newman black holes with different angular momentum and charge parameters, keeping the sum $(Q^2+J^2/M^2)$ fixed. For extremal Kerr-Newman black holes satisfying (\ref{extremalalpha}), there exists a narrow range of frequencies $\omega_{\rm{sl}}<\omega<\omega_{\rm{max-ex}}$ that can lead to overspinning. However, this can easily be fixed by employing backreaction effects. The induced increase in the angular momentum of the event horizon can be calculated as
\begin{equation}
\Delta \omega = \frac{\delta J}{4M^3}=\frac{1}{M}\frac{(1+\zeta \xi)\sqrt{2\zeta \xi +\alpha^2}-\alpha}{4\xi}
\label{probdeltaomegeext}
\end{equation}
where we have used 
\[
\delta J=\frac{m}{\omega} \delta M=\left( \frac{1}{\omega_{\rm{max-ex}}}\right) M\zeta
\]
As we have argued previously it is critical to calculate $\Delta \omega$ for the upper limit $\omega_{\rm{max-ex}}$. Substituting the superradiance limit for extremal black holes and the induced increase in the superradiance limit derived in (\ref{probdeltaomegeext}), the modified value of the superradiance limit takes the form:
\begin{eqnarray}
\omega'_{\rm{sl}}&=& \omega_{\rm{sl}}+\Delta \omega \nonumber  \\
&=&\frac{1}{M\left[\frac{1}{\alpha}+\alpha\right]}+ \frac{1}{M}\frac{(1+\zeta \xi)\sqrt{2\zeta \xi +\alpha^2}-\alpha}{4\xi}
\end{eqnarray}
To overspin an extremal Kerr-Newman black hole, the frequency of a test field should be less than the maximum value derived in (\ref{omegamaxprobex}). A test field with such a low frequency can have a positive absorption probability only for a class of extremal black holes satisfying (\ref{extremalalpha}). We checked if the overspinning can be fixed by the induced increase in the angular momentum of the black hole, which modifies the superradiance limit. Due to the induced increase in the angular momentum, the superradiance limit increases. The absorption of the modes with frequencies lower than the modified value of the superradiance limit is prevented, though their absorption probability appears to be positive when one ignores backreaction effects based on the induced increase in the angular momentum. If this modified value exceeds the maximum value derived in (\ref{omegamaxprobex}), the absorption of all the challenging modes with low frequencies and positive absorption probabilities will be prevented.
One observes that the modified value of the superradiance limit exceeds the maximum value of the frequency that can lead to overspinning, for any value of $\alpha$ in the range $0<\alpha<1$. To clarify this, we have plotted $\omega_{\rm{max-ex}}$, $\omega_{\rm{sl}}$, and $\omega_{\rm{sl}}'$ as a function of $\alpha$ in figure (\ref{figure}). 
The fact that the modified value of the superradiance limit exceeds the upper limit of frequency $\omega_{\rm{max-ex}}$, indicates that the absorption of the fine-tuned frequencies in the narrow range $\omega_{\rm{sl}}<\omega<\omega_{\rm{max-ex}}$ will be prevented; i.e. the event horizon cannot be destroyed. Taking absorption probabilities into consideration  fixes the overspinning problem for the extremal case as well as the nearly extremal case.

\section{Summary and conclusions}
In this work we have re-considered the overspinning problem for Kerr-Newman black holes. First, we have scrutinized the recent analysis by Sorce and Wald where they employ a variational method and expand the field configurations to second order in the small parameter $\lambda$.  Sorce and Wald claim that they have obtained an expression for the full second order correction $\delta^2 M$ without having to calculate the backreaction effects explicitly. In a recent paper we have also employed the method developed by Sorce and Wald for MTZ black holes. In that work we have imposed that $\delta M$ is a first order quantity itself, for test bodies and fields. We have noticed that imposing this non-controversial fact leads to order of magnitude problems in the SW method. Therefore we have concluded that it would be better to calculate the backreaction effects explicitly. Here we argued that the order of magnitude problems do not pertain to the case of MTZ black holes, they also appear in the case of Kerr-Newman black holes. The argument is simple. Since $\delta M$ is inherently a first order quantity:
\[
\lambda \delta M \Rightarrow \mbox{second order, not first}
\]
\[
\lambda \epsilon \delta M \Rightarrow \mbox{third order, not second}
\]
\[
\lambda^2 (\delta M)^2 \Rightarrow \mbox{fourth order, not second}
\]
\[
\lambda^2 \delta^2 M \Rightarrow \mbox{fourth order, not second}
\]
Sorce and Wald claim that the function $f(\lambda)$  in the equation (\ref{swmain}) --which is the equation (119) in \cite{w2}-- can be made negative for the terms first order in $\lambda$ and the contribution of the terms second order in $\lambda$ makes $f(\lambda)$ positive again. This implies that the previous results due to Hubeny \cite{hu}, Jacobson-Sotiriou \cite{js}, and D\"{u}zta\c{s}-Semiz \cite{overspin} are reproduced. Nearly extremal black holes can be destroyed by test bodies and fields and the destruction of the event horizon is fixed by employing backreaction effects. We showed that these correct results cannot be reproduced by SW method. The leading term in $f(\lambda)$ is of the second order $(M^2 \epsilon^2)$, whereas the contribution of the second order perturbations are of the fourth order $(\lambda^2 \delta^2 M)$. 
Therefore the effect of second order corrections cannot be incorporated using SW method, unless one fallaciously imposes $\delta M \sim M$. This assumption apparently contradicts the test body/field approximation. For that reason, backreactions should be identified and explicitly calculated for every specific problem.

Based on the argument that the SW method is invalid, we re-visited the overspinning problem for extremal and nearly extremal Kerr-Newman black holes. In a recent paper we had shown that there exists a class of extremal Kerr-Newman black holes which can be overspun by neutral test fields \cite{generic}. The overspinning is possible if the extremal Kerr-Newman black hole satisfies $\alpha^2 \equiv J^2/M^4 <1/3$. (The dimensionless parameter $\alpha$ is introduced to distinguish black holes with different angular momentum and charge parameters keeping the sum $(Q^2+J^2/M^2)$ fixed. See equations (\ref{paramalpha}) and (\ref{mainextalpha}). ) There, we gave numerical examples and compared our results with previous claims. Here we showed that nearly extremal Kerr-Newman black holes can also be overspun independent of the value of $\alpha$. In our analysis we do not ignore the contribution of the second order terms $(\delta M)^2$ and $(\delta J)^2$ which drastically changes the results. Therefore we need to calculate the backreaction effects to complete our analysis. In this work, we employed the backreaction effects due to the increase in the angular velocity of the horizon which was first suggested by Will \cite{will}. The induced increase in the angular velocity of the event horizon leads to an increase in the superradiance limit. This prevents the absorption of the modes that could potentially overspin the black hole. We showed that this backreaction brings further restrictions to the classes of extremal and nearly extremal black holes that can be overspun by test fields. Overspinning is prevented for nearly extremal black holes with $\alpha \gtrsim 0.50$ and for extremal black holes with $\alpha \gtrsim 0.31$. We noted that the effect of backreactions is an open problem and there could be different sorts of backreaction effects that could potentially bring a full solution to the overspinning problem. 

The results derived in this work can also be exploited to evaluate the possibility to overspin Kerr black holes. In the limit $Q \to 0$, extremal Kerr black holes are identified with $\alpha\equiv J/M^2=1$ whereas nearly extremal ones are parametrised as $\alpha^2=1-\epsilon^2$. There is no overspinning problem for extremal Kerr black holes even if one ignores the backreaction effects, since $\alpha^2=1>(1/3)$. The backreaction effects due to the increase in the superradiance limit fixes the overspinning problem for nearly extremal Kerr black holes, since $\alpha=1-\epsilon^2>(0.50)$. These findings are in accord with previous results on Kerr black holes.

In all the previous thought experiments --including the works of this author-- the absorption probability of test fields was ignored. Ignoring the probability corresponds to assuming that it is of the order of unity. However, the interaction of black holes with test fields is a scattering problem. A fraction of the test field is absorbed by the black hole, while part of it is reflected back to infinity. In section (\ref{prob}), we have incorporated the absorption probabilities in the thought experiments to test whether the event horizon can be destroyed. The fact that only a small fraction of the challenging modes is absorbed by the black holes, fundamentally changes the course of the analysis in favour of the cosmic censorship conjecture. We calculated the absorption probability for test fields with frequency close to the superradiance limit, using the seminal results by Page \cite{page}. The absorption probability of a test field with frequency $\omega=m\Omega(1+\xi)$ turns out to be of the order $O(\xi)$. The probability approaches zero for optimal perturbations with frequency $\omega=m\Omega$, which implies that these fields are entirely reflected back to infinity. The parameters of the space-time remain invariant after the interaction with  these test fields. Hence, the event horizon cannot be destroyed.  For the  nearly extremal case, we derived that there exists a class of Kerr-Newman black holes identified by $\alpha \lesssim 0.01401$, which can be destroyed by test fields. Overspinning occurs due to the fact that test fields with very low frequencies can be absorbed by these black holes. The contribution of these test fields to the angular momentum parameter will be large, since it is inversely proportional to the frequency. However the induced increase in the angular momentum of the event horizon is also large for these perturbations. Therefore the overspinning is easily fixed by employing the backreaction effects. We noted that one can also argue that these fields cannot be treated in the test field approximation due to the large magnitude of $\delta J$. This argument would render the overspinning problem irrelevant. For the case of extremal black holes we derived that there exists a set of fine-tuned parameters that can overspin Kerr-Newman black holes with $\alpha \lesssim 0.70707$. However, the range of frequencies is very narrow and the overspinning problem is fixed by employing backreaction effects. Both for extremal and nearly extremal Kerr-Newman black holes, the ultimate solution to the overspinning problem follows by the incorporation of the absorption probabilities into the analysis.

In a recent paper we argued that fermionic fields lead to a generic destruction of the event horizon in the classical picture \cite{generic}. Since the fermionic fields do not obey the weak energy condition, one cannot find a lower bound for the energy of a fermionic field similar to the condition (\ref{needham}). The absorption probability is positive definite and it approaches zero only as $\omega$ approaches zero, as confirmed by Page \cite{page}. For that reason, the arguments about the absorption probability of scalar fields and its effect on the overspinning problem developed in this paper, do not apply to fermionic fields.

%
%

\end{document}